\documentclass[letterpaper, 10 pt, conference]{ieeeconf}  
\IEEEoverridecommandlockouts
\usepackage{cite}
\usepackage{amsmath,amssymb,amsfonts}
\usepackage{graphicx}
\usepackage{textcomp}
\usepackage{xcolor}
\usepackage{accents}
\usepackage[mathscr]{euscript}
\usepackage{mathrsfs}
\usepackage{mathtools}
\usepackage{textcomp, gensymb}
\usepackage{booktabs}
\usepackage{multirow}
\usepackage{array}
\usepackage{algorithm, algpseudocode}
\usepackage{subcaption}
\usepackage[capitalize]{cleveref}

\begin{document}

\title{Communication-Based Distributed Control of Large-Scale District Heating Networks}

\author{Audrey~Blizard,~and~Stephanie~Stockar%
\thanks{This material is based upon work supported by the National Science Foundation Graduate Research Fellowship Program under Grant No. DGE-1343012}%
\thanks{Audrey Blizard (blizard.1@osu.edu) and Stephanie Stockar (stockar.1@osu.edu) are with the  Department of Mechanical and Aerospace Engineering and the Center for Automotive Research, The Ohio State University, 930 Kinnear Road, Columbus, OH 43212 USA}}

\maketitle

\begin{abstract}
This paper presents a non-cooperative distributed model predictive controller for the control of large-scale District Heating Networks. To enable the design of this controller a novel information passing scheme and feasibility restoration method are created, allowing the local controllers to achieve a global consensus while minimizing a local cost function. The effectiveness of this controller is demonstrated on an 18-user District Heating Network decomposed into six subsystems. The results show that the developed control scheme effectively uses flexibility to manage the buildings' heat demands reducing the total losses by 14\% and the return temperature by 37\%.
\end{abstract}

\section{Introduction}
District heating networks (DHNs) are a promising method to decarbonize the energy-intensive process of heating buildings and allow for the integration of the growing number of distributed energy resources \cite{vandermeulenControllingDistrictHeating2018}. They do this by allowing heat generated in distributed locations to be transported through a network of underground pipes to non-local buildings. The heat is then extracted by the buildings via heat exchangers, and the cooled water is returned to the distributed heat sources creating a closed loop.\par
There is a large potential for increasing the efficiency of these networks through smart control strategies. One of the main sources of control authority in these networks is the flexibility provided by the thermal capacity of the network and connected users \cite{vandermeulenControllingDistrictHeating2018}. However, the large-scale nature of DHNs makes effectively utilizing this flexibility extremely challenging. One solution to quantifying and using this flexibility is through reduced order modeling where the flexibility of individual buildings is aggregated into neighborhoods. This method showed a 16\% reduction in peak demands \cite{salettiEnablingSmartControl2021}. Another work demonstrates that when implemented on a real-world DHN, this flexibility can reduce peak load energy by 34\% \cite{vanoevelenTestingEvaluationSmart2023}. However, this study achieved these results only considering the active control of a single building in the multi-building network due to the challenges of large-scale control, indicating the need for an effective control strategy.\par
In previous work, the authors have shown the potential of considering flexibility at a building level to reduce wasted flow through the individual control of buildings' flow supplies \cite{blizardUsingFlexibilityEnvelopes2024b}. This work used a hierarchical controller to solve the large-scale optimization problem, which solves the local optimization problem over a series of discrete pressure drops. While able to reduce overall wasted flow by 67\%, implementing this controller requires computing many discrete pressure solutions which can be difficult and time-consuming. The work presented here looks to reduce computational demands by using distributed rather than hierarchical model predictive control (MPC) to get similar performance improvements without unnecessary solution calculations.\par
The challenge with developing a distributed MPC (dMPC) for DHNs comes from the nonlinearity in the algebraic coupling constraints and the bilinearity in the system dynamics, resulting in a non-convex optimization problem with system-wide equality constraints. Due to these characteristics, generalized dMPC frameworks are not effective for this problem \cite{mullerEconomicDistributedModel2017}. Additionally, distributed optimization methods, such as ADMM \cite{rostamiADMMbasedDistributedModel2017}, rely on convexity to ensure problem convergence. Even distributed optimization methods designed for non-convex problems, such as ALADIN \cite{houskaAugmentedLagrangianBased2016}, require decoupled constraints, which is not possible in considering the pressure balancing constraints in the network. Additionally, case-specific schemes were considered, including those used in building HVAC systems \cite{yaoStateArtReview2021} and microgrids \cite{huModelPredictiveControl2021}, but none were able to compensate for the unique constraints and dynamics seen in large-scale DHNs.\par
Instead, in this paper, a custom communication-based distributed MPC framework, with a specialized information passing scheme and feasibility recovery mechanism is developed that meets the unique requirements of this problem. Through this framework, the subsystems converge to a Nash equilibrium, ensuring agreement between neighboring subsystems. The developed dMPC is demonstrated in the loss-minimizing control of an 18-user DHNs.\par

\section{Problem Formulation}
This section presents the network model and centralized problem formulation that will be approximated by the distributed controller.
\subsection{Network Description}
A directed graph, $\mathcal{G} = (\mathcal{V},\mathcal{E})$ will be used to represent the heat supply network of the DHN. The edges of the graph $\mathcal{E}$ represent the pipes in the network, while the nodes $\mathcal{V}$ are the interconnections of these pipes. These interconnections are captured in the graph's incidence matrix, $\Lambda$. There are four unique edge types identifiable in a DHN, the feeding edges $\mathcal{E}_F$, the return edges $\mathcal{E}_R$, the user edges $\mathcal{E}_U$, and the bypass edges $\mathcal{E}_{By}$. The bypass edges connect the feeding and return networks, passing flow directly without heat being extracted by a user. Note that $\mathcal{V}_F =\operatorname{innodes}\left(\mathcal{E}_F\right)\cup\operatorname{outnodes}\left(\mathcal{E}_F\right)$ are the feeding nodes and $\mathcal{V}_R =\operatorname{innodes}\left(\mathcal{E}_R\right)\cup\operatorname{outnodes}\left(\mathcal{E}_R\right)$ are the return nodes. Additionally, the set of non-user edges is denoted as $\mathcal{E}_N = \mathcal{E}_F\cup\mathcal{E}_R\cup\mathcal{E}_{By}$. A sample DHN demonstrating these edge sets is presented in \cref{fig:graphs}.\par
Each edge in $\mathcal{E}_N$ has an associated friction coefficient $\zeta_u$, volume $V$, heat transfer coefficient $hA_s$, flow rate $\dot{m}$, and inlet temperature $T_{in}$. The temperature $T$ of one of these edges can be modeled using a well-mixed approach as
\begin{gather}
    \label{eq:Tpipe}
    \frac{d}{dt}T=c_1\cdot\dot{m}\cdot T_{in} + c_2 T_{amb} - (c_1\cdot\dot{m}+c_2) T\\
    {c_1} = \frac{1}{\rho V},\quad {c_2} = \frac{hA_s}{\rho c_pV}
\end{gather}
where $\rho$, $c_p$ are the density and specific heat capacity of the operating fluid in the network, respectively. Additionally, $T_{amb}$ is the ground temperature, used to calculate losses from the pipe.\par
In the user edges, the friction coefficient of the edge is variable, allowing the user to control the mass flow rate in these edges. This variable friction coefficient is given by 
\begin{equation}\label{eq:valve}
    \zeta_U = \mu\left(\frac{1}{\theta}-1\right)^2
\end{equation}
determined by the valve position $\theta_{min}<\theta<1$, where $\theta_{min}$ is associated with some minimum pressure loss and $\mu$ is a scaling coefficient. This valve model allows the optimization problem to be better conditioned. Additionally, the bulk temperatures of these edges are modeled as a constant $T_{setR}$, as the dynamics of the heat exchanger are much faster than those of the other network pipes.\par
The heating plant will be modeled as a root and terminal node of the graph $v_{0^+}, v_{0^-}$. Any outedges of $v_{0^+}$ will receive water with a predefined temperature $T_0$. The flow rate into the network, modeled as a nonzero node flow, $\dot{m}_{0}$ is a control variable.

\subsection{Centralized Optimization}
The centralized optimization problem being considered in this paper, originally proposed in \cite{blizardUsingFlexibilityEnvelopes2024b} is given by
\begin{subequations}
    \begin{equation}\label{eq:opt_cost} 
       \min_{\dot{m}_0,\theta_{min}<\theta<1} f(\dot{m}_{\mathcal{E}},T_{{\mathcal{E}_N}},SOE_U) \quad \text{subject to:} 
    \end{equation}
    \begin{equation}\label{eq:opt_temp}
        \frac{d}{dt}T_{{\mathcal{E}_N}} = A(\dot{m}_{{\mathcal{E}_N}})T_{{\mathcal{E}_N}} + B \begin{bmatrix}T_0\\T_{setR}\\T_{amb}\end{bmatrix}
    \end{equation}
    \begin{equation}\label{eq:opt_qp}
        \dot{Q}_p = \dot{m}_{{\mathcal{E}}_U}c_p(T_{in\ {\mathcal{E}}_U}-T_{setR})
    \end{equation}
    \begin{equation}\label{eq:opt_soe}
        \frac{d}{dt}SOE_U = \dot{Q}_p(k)-\dot{Q}_{out}
    \end{equation}
    \begin{equation}\label{eq:opt_flex}
         -C\Delta T_b\leq SOE_U\leq C\Delta T_b
    \end{equation}
    \begin{equation}\label{eq:opt_dP}
        \Delta P_{\mathcal{E}} = \zeta_{\mathcal{E}} \dot{m}_{\mathcal{E}}^2,\quad \zeta_{U}\subset\zeta_{\mathcal{E}}
    \end{equation}
    \begin{equation}\label{eq:opt_flow}
        \Lambda \dot{m}_{\mathcal{E}} = \dot{m}_{\mathcal{V}}
    \end{equation}
    \begin{equation}\label{eq:opt_pressure}
    \Delta P_{\mathcal{E}} = \Lambda^T P_{\mathcal{V}},\quad P_{\mathcal{V}}(v_{0^-})=0
    \end{equation}
\end{subequations}
The cost to be minimized, \cref{eq:opt_cost}, can be any function of the network variables described below, where the two control variables are the plant mass flow and the valve positions.\par
\Cref{eq:opt_temp} represents the temperature dynamics of the non-user edges in the network, assembled from \cref{eq:Tpipe}. The $A$ matrix represents the interconnections between the pipes and includes $c_1$, $c_2$, and the mass flow rate in these edges. The $B$ matrix considers edges connected to both the plant and the outlet of the user edges and accounts for losses to the environment. The full details on the creation of the $A$ and $B$ matrices are found in \cite{blizardGraphBasedTechniqueAutomated2024}.\par
\Cref{eq:opt_qp} calculate the heat delivered to the users, $\dot{Q}_p$, using the temperature of the edges entering the heat exchangers $T_{in\ \mathcal{E}_U}$. Any heat delivered above or below the nominal demand of the building, $\dot{Q}_{out}$, contributes to changing its state of energy $SOE_U$ given in \cref{eq:opt_soe}. This state of energy is constrained to remain within a flexibility envelope \cite{reyndersGenericCharacterizationMethod2017}, determined by the building's heat capacity $C$ and acceptable temperature deviation $\Delta T_{b}$ as given in \cref{eq:opt_flex}.\par
Algebraic constraints are used to calculate the mass flow rates. \Cref{eq:opt_dP} gives the pressure losses in every edge. The pressure balance in the network is enforced using \cref{eq:opt_pressure}, while the mass flow rate in the edges is determined by \cref{eq:opt_flow}, where the vector of node flows $\dot{m}_{\mathcal{V}}$ is 
\begin{equation}
    \dot{m}_{\mathcal{V}}(v) =\begin{cases}
    \dot{m}_0 & \text{if } v = v_{0^+}\\
    -\dot{m}_0 & \text{if } v = v_{0^-}\\
    0 &\text{otherwise}
    \end{cases}
\end{equation}
\begin{figure}
    \centering
    \includegraphics[width=1\linewidth]{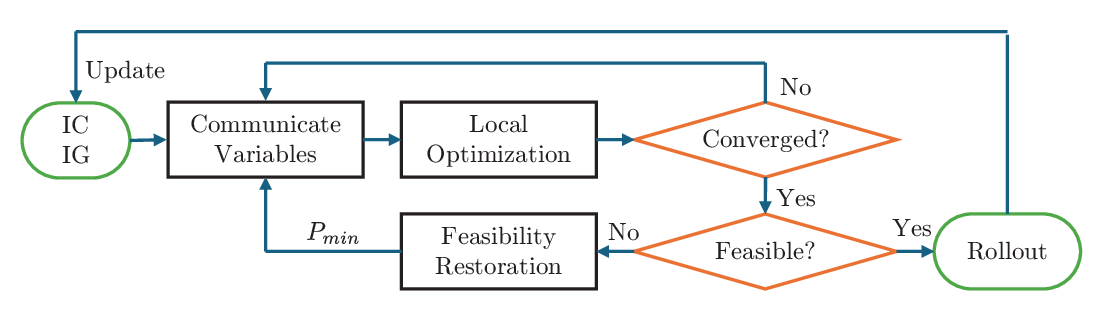}
    \caption{Flowchart of steps taken in the presented dMPC scheme.}
    \label{fig:flowchart}
\end{figure}
\section{Distributed Control Architecture}
In this section, a non-cooperative communication-based dMPC is developed. Distributed controllers will solve local optimization problems, based on the predicted control trajectories of the neighboring subsystems. Then, the solutions to these problems will be used to update the information passed to neighboring subsystems. This process will continue until all subsystems have converged to a local solution, which is aggregated and implemented. A flowchart of the steps taken by this controller is presented in \cref{fig:flowchart}. As a non-cooperative controller, this solution will converge to a Nash Equilibrium where no subsystem can unilaterally improve its control performance \cite{christofidesDistributedModelPredictive2013}, and can only approximate the solution of the centralized problem. A case study is used to demonstrate the effectiveness of this approximation. 

\begin{figure}
    \centering
    \includegraphics[width=1\linewidth,trim={75 15 50 10},clip]{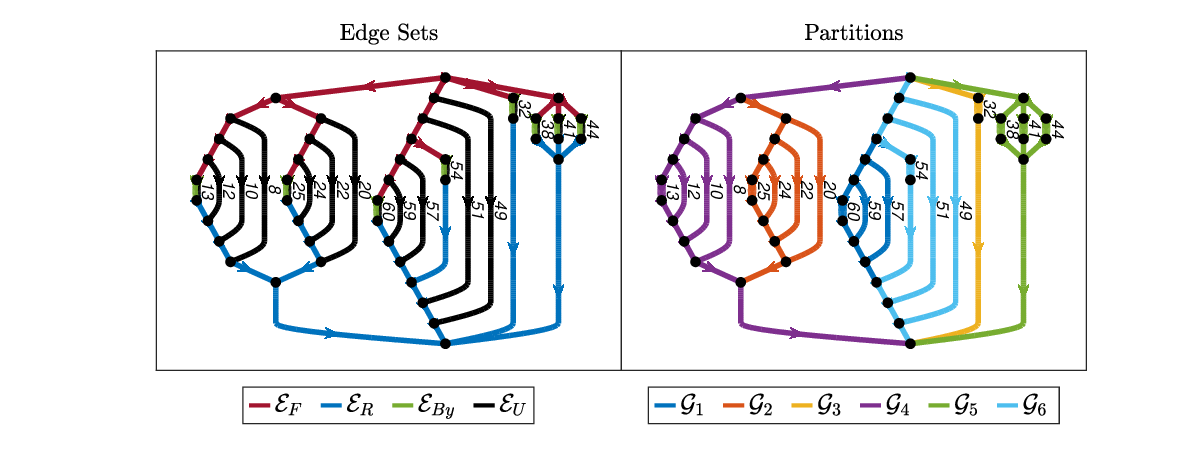}
    \caption{Graphs of case study network with user edges numbered.}
    \label{fig:graphs}
\end{figure}

\subsection{Network Partitioning}
The first step to creating the distributed controller is partitioning the network into smaller controllable subsystems, each represented by a local subgraph $\mathcal{G}_i = (\mathcal{V}_i,\mathcal{E}_i)$, $i=1\dots n_g$. To create these subgraphs the original set of edges $\mathcal{E}$ is divided into unique, non-overlapping sets. The local node sets are defined as $\mathcal{V}_i = \operatorname{innodes}\left(\mathcal{E}_i\right)\cup\operatorname{outnodes}\left(\mathcal{E}_i\right)$. A partitioning will only happen downstream of a feeding node as $\operatorname{indegree}(v_F) = 1$ and upstream of a return node as $\operatorname{outdegree}(v_R) = 1$.  In each of the subgraphs, the set of local root and terminal nodes are identified as $\mathcal{V}_{i^+}$ and $\mathcal{V}_{i^-}$. Additionally, these root and terminal nodes will have a matching node with inverse edge type in another subgraph, indicating where a cut was made. These matching node sets are denoted as $\mathcal{V}_{j^o}$ and $\mathcal{V}_{j^e}$ respectively, where $j$ does not have to be a single subgraph. Handling the behavior at the overlapping nodes is the key step to the distributed control design. The method presented is applicable to partitions with $\operatorname{card}\left(\mathcal{V}_{i^+}\right)>1$ and $\operatorname{card}\left(\mathcal{V}_{i^-}\right)>1$, so there are no constraints on how these partitions are chosen.
The six partitions chosen for the test case are presented in \cref{fig:graphs}.

\subsection{Variable Communication}
Three variable sets: network temperatures, node pressures, and node flow rates, must be passed between subsystems for the local solutions to be consistent with the global network behavior. All passed variables are denoted by $\left(\cdot\right)^{p}$. Each type of variable has a different passing direction to give subsystems at least one degree of freedom in their local optimization problem. Temperature is passed consistent with the flow direction according to 
\begin{equation}
\begin{gathered}
    v_{j^o}\rightarrow v_{i^+}\quad \forall\ v\in\mathcal{V}_F\\
    v_{i^-}\rightarrow v_{j^e}\quad \forall\ v\in\mathcal{V}_R
\end{gathered}
\end{equation}
The node pressures are passed by the downstream subsystem if the node is a return node and by the upstream subsystem if the node is a feeding node, according to 
\begin{equation}
    \begin{gathered}
        v_{j^o}\rightarrow v_{i^+}\quad \forall\ v\in\mathcal{V}_F\\
        v_{j^e}\rightarrow v_{i^-}\quad \forall\ v\in\mathcal{V}_R
    \end{gathered}
\end{equation}
This passing scheme is chosen as $\operatorname{outdegree}\mathcal{V}_R=1$ and $\operatorname{indegree}\mathcal{V}_F=1$, meaning a single subsystem will determine the node pressure. The plant pressure, while technically a feeding node, must be determined by a downstream subsystem. The subsystem with the largest total downstream heat demand is selected to determine the plant pressure, as this subsystem will, on average, require more flow, and higher flow requires a larger total pressure drop. \par
\begin{figure}
    \centering
    \includegraphics[width=.9\linewidth,trim={135 60 75 40},clip]{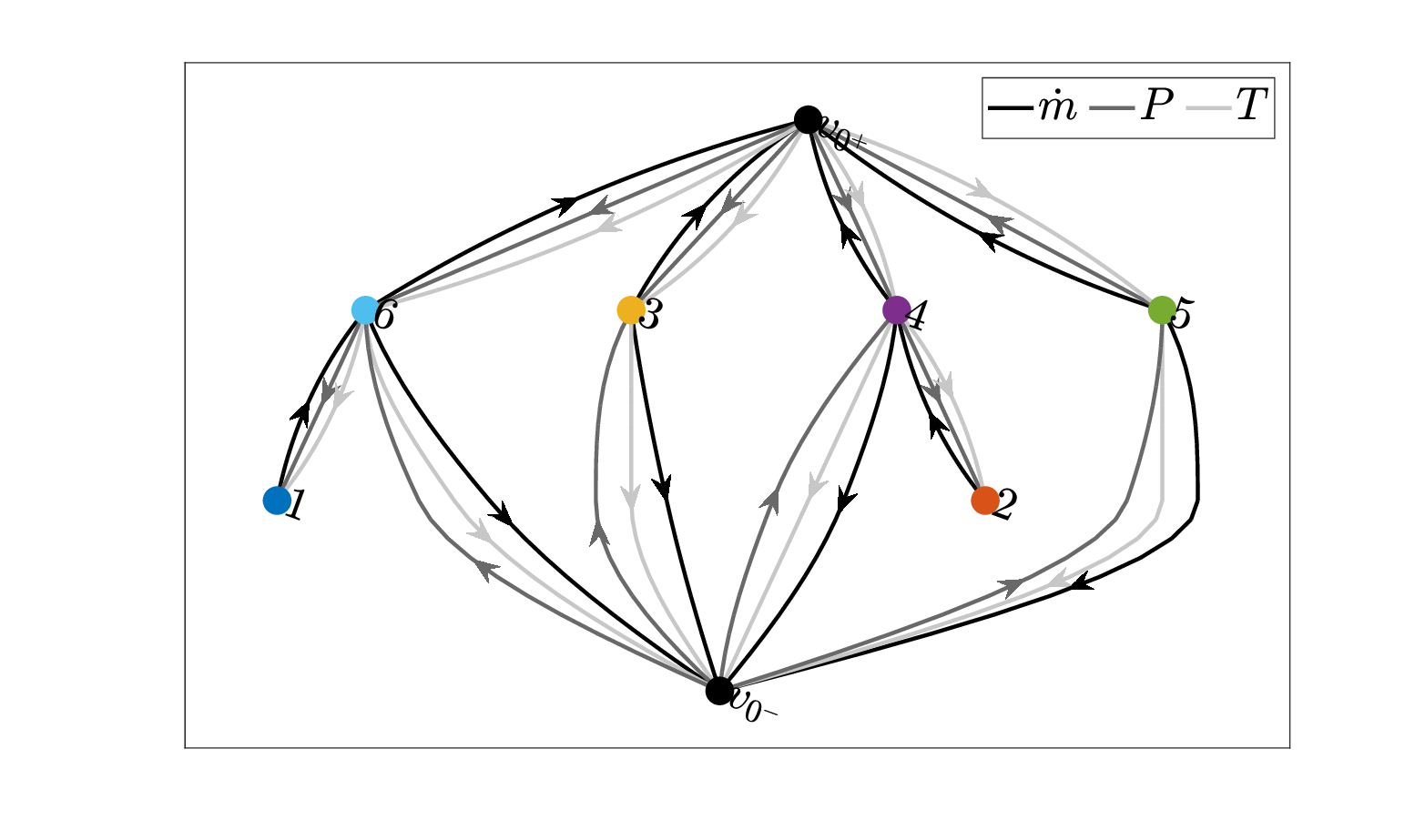}
    \caption{Communication graph between subsystems.}
    \label{fig:comm}
\end{figure}

 
The passing of the mass flow rate is the inverse of pressure passing, given by 
\begin{equation}
    \begin{gathered}
        v_{i^+}\rightarrow v_{j^o}\quad \forall\ v\in\mathcal{V}_F\\
        v_{i^-}\rightarrow v_{j^e}\quad \forall\ v\in\mathcal{V}_R
    \end{gathered}
\end{equation}
This information passing scheme is chosen because, in cases where one subsystem has both an upstream and downstream connection to the same neighboring subsystem, it allows the encompassed subsystem to choose any mass flow rate as long as it meets the required pressure drop. The authority over both the inlet and outlet mass flow (which must be equal) allows for the degree of freedom necessary to meet heat demands and signal to the encompassing subsystem its planned control action. Additionally, in cases where multiple subsystems are leaving the same feeding node, from a full network perspective, the mass flow split will equalize the pressure loss between the downstream edges. The network split will match the selected control trajectory as all downstream subsystems are constrained to have the same node pressure when determining mass flow rate. The passing graph of the partitions chosen for the case study is presented in \cref{fig:comm}.
\subsection{Distributed Optimization Problem}
The local optimization problem solved by each subnetwork is given by
\begin{subequations}
    \begin{equation}\label{eq:i_cost} 
       c_i = \min_{\widetilde{\dot{m}}_i,\theta_{min}<\theta_i<1} f(\dot{m}_{\mathcal{E}_i},T_{{\mathcal{E}_N}_i},SOE_{U_i}) \quad \text{subject to:}
    \end{equation}
    \begin{equation}\label{eq:i_temp}
        \frac{d}{dt}T_{{\mathcal{E}_N}_i} = A(\dot{m}_{{\mathcal{E}_N}_i})T_{{\mathcal{E}_N}_i} + B \begin{bmatrix}T_{0_i}\\T_{setR}\\T_{amb}\end{bmatrix}
    \end{equation}
    \begin{equation}\label{eq:i_qp}
        \dot{Q}_p = \dot{m}_{{\mathcal{E}}_{U_i}}c_p(T_{in\ {\mathcal{E}}_{U_i}}-T_{setR})
    \end{equation}
    \begin{equation}\label{eq:i_soe}
        \frac{d}{dt}SOE_{U_i} = \dot{Q}_p(k)-{\dot{Q}_{out}}_i
    \end{equation}
    \begin{equation}\label{eq:i_flex}
         -C_i\Delta {T_b}\leq SOE_{U_i}\leq C_i\Delta {T_b}
    \end{equation}
    \begin{equation}\label{eq:i_dP}
        \Delta P_{\mathcal{E}} = \zeta_{\mathcal{E}} \dot{m}_{\mathcal{E}}^2,\quad \zeta_{U}\subset\zeta_{\mathcal{E}}
    \end{equation}
    \begin{equation}\label{eq:i_flow}
        \Lambda_i \dot{m}_{\mathcal{E}_i} = \dot{m}_{\mathcal{V}_i}
    \end{equation}
    \begin{equation}\label{eq:i_pressure}
    \Delta P_{\mathcal{E}_i} = \Lambda_i^T P_{\mathcal{V}_i},\quad P_{\mathcal{V}_i}(v_{0^-})=0
    \end{equation}
\end{subequations}
where \cref{eq:i_dP,eq:i_flex,eq:i_soe,eq:i_qp} are unchanged, except for considering only local variables. In \cref{eq:i_temp}, ${T_0}_i$ is the local inlet temperatures given by 
\begin{equation}
    {T_0}_i(e) = \begin{cases}
        T_0 & \text{if } \operatorname{innode}\left(e\right)=v_{0^+}\\
        T_j^p(e_j) & \text{if } \operatorname{innode}\left(e\right)\in\mathcal{V}_{i^+}
    \end{cases}
\end{equation}
where $e$ are the local nonuser edges connected to the plant or other subsystems, and $e_j$ is the edge where $\operatorname{outnode}\left(e_j\right) = \operatorname{innode}\left(e\right)$. In subnetworks where any node pressures are dictated by a different subsystem, additional constraints
\begin{equation}\label{eq:constrP}
    P_{\mathcal{V}_i}(v) = P^{p}_{\mathcal{V}_j}(v),\quad v = \left(\mathcal{V}_{i^+}\backslash\mathcal{V}_F\right) \cup \left(\mathcal{V}_{i^-}\backslash\mathcal{V}_R\right)
\end{equation}
are added to the local optimization problem. The vector of node mass flow rates in \cref{eq:i_flow}, $\dot{m}_{\mathcal{V}_i}$, is now given by 
\begin{equation}
    \dot{m}_{\mathcal{V}_i}(v) =\begin{cases}
    \dot{m}_{e_i}(v) & \text{if } v = v_{0^+}\\
    \dot{m}_{o_i}(v) & \text{if } v = v_{0^-}\\
    \dot{m}_{e_i}(v) & \text{if } v\in\mathcal{V}_F\ \&\ v\in \mathcal{V}_{i^+}\\
     \dot{m}_{o_i}(v) & \text{if } v\in\mathcal{V}_R\ \&\ v \in \mathcal{V}_{i^-}\\
     \dot{m}_{e_j}^{p}(v) & \text{if } v\in\mathcal{V}_R\ \&\ v\in\mathcal{V}_{i^e}\\
     \dot{m}_{o_j}^{p}(v)& \text{if } v\in\mathcal{V}_F\ \&\ v\in\mathcal{V}_{i^o}\\
    0 &\text{otherwise}
    \end{cases}
\end{equation}
where $\dot{m}_{e_i}\geq0$ is a vector of controllable inlet flows and $\dot{m}_{o_i}\leq0$ is a vector of controllable outlet flows. Additionally, these mass flow rate variables are combined into the control variable as
\begin{equation}
    \widetilde{\dot{m}}_i = \begin{bmatrix}
        \dot{m}_{e_i}\\
        \dot{m}_{o_i}
    \end{bmatrix}
\end{equation}
Finally, this information is used to recover the centralized plant mass flow $\dot{m}_0$ as
\begin{equation}
    \dot{m}_0 = \sum_{i=1\dots n_g\text{ s.t. } v_{0^+}\in\mathcal{V}_i} \dot{m}_{e_i}^{p}(v_{0^+}) 
\end{equation}
\subsection{Convergence}
After each subsystem finds its local control trajectories based on the predicted trajectories from neighboring subsystems, the variables to be passed,
defined as 
\begin{equation}
    v_{pass} = \begin{bmatrix}
        T_i(e_{out})\\
        P_{\mathcal{V}_i}\left(\mathcal{V}_{i^e}\backslash\mathcal{V}_F \right)\\
        P_{\mathcal{V}_i}\left(\mathcal{V}_{i^o}\backslash\mathcal{V}_R \right)\\
        \dot{m}_{e_i}(\mathcal{V}_{i^+}\backslash v_{0^+})\\
        \dot{m}_{o_i}(\mathcal{V}_{i^-}\backslash v_{0^+})
    \end{bmatrix}
\end{equation}
are updated according to
\begin{equation}
    v_{pass}^{p_{new}} = \omega v_{pass}^p + (1-\omega) v_{pass}
\end{equation}
where $\omega$ is the step size.
A subsystem is considered converged when 1) the change in all of the variables passed by the subsystem and the subsystem cost meet
\begin{equation}
    \begin{bmatrix} v_{pass}^{p}-v_{pass}\\ c_i^p-c_i
    \end{bmatrix} \leq \epsilon
\end{equation}
where $\epsilon$ is a vector threshold values, and 2) when all subsystems passing variables into the subsystem have also converged. As different subsystems have different degrees of interaction, subsystems with low or no interaction converge first, allowing dependent subsystems to follow. 
\subsection{Feasibility Restoration}
 The assumptions made in the centralized problem formulation ensure that in the centralized case, a feasible solution is always available.\\
 \emph{Feasibility Assumptions:}
\begin{enumerate}
    \item The flexibility envelope of an individual building is constant or increasing in size.
    \item It is always possible to meet a building's heat demand due to the algebraic model of the heat exchanger and unconstrained plant flow.
    \item The only way heat is added to a building is through the DHN. Solar irradiation and ambient temperatures exceeding the building temperature are not considered.
\end{enumerate}
From these assumptions, it is evident that, in the centralized case, if the building's heat capacity starts within the flexibility envelope, it will be able to remain within the flexibility envelope. However, in the distributed case, a single subsystem dictates the pressure at the plant, effectively limiting the supply flow into any of the other subsystems. This limit can mean that the required heat is not available to prevent a building from exceeding the lower limit of its flexibility envelope, removing assumption 2. To prevent this, a feasibility recovery algorithm is implemented, which allows an initially unfeasible subsystem to temporarily control the plant pressure. The cost function solved by infeasible subsystems is switched to
\begin{equation}
     \min_{\widetilde{\dot{m}},\theta_{min}<\theta_i<1} P_{slack}
\end{equation}
where $P_{slack}$ is the increase from the dictated plant pressure, and the constraint established in \cref{eq:constrP} is relaxed to 
\begin{equation}
    P_{\mathcal{V}_i}(v_{0^+}) = P^{p}_{\mathcal{V}_j}(v_{0^+})+P_{slack}
\end{equation}
This new optimization problem finds the minimum plant pressure required by the subsystem to maintain feasibility. The minimum plant pressure, $P_{min}=P^{p}_{\mathcal{V}_j}(v_{0^+})+P_{slack}$, is transmitted to the plant pressure control subsystem, which is then used as a lower limit for the plant pressure adding the local constraint
\begin{equation}
    P_{\mathcal{V}_i}(v_{0^+})\geq P_{min}
\end{equation}
Then, the initially infeasible subsystem returns to its original cost function, and the regular optimization procedure resumes. By implementing this feasibility restoration, the overall controller guarantees feasibility in the distributed case. 
\section{Results}
\begin{figure}
    \centering
    \includegraphics[width=1\linewidth,trim={40 50 49 35},clip]{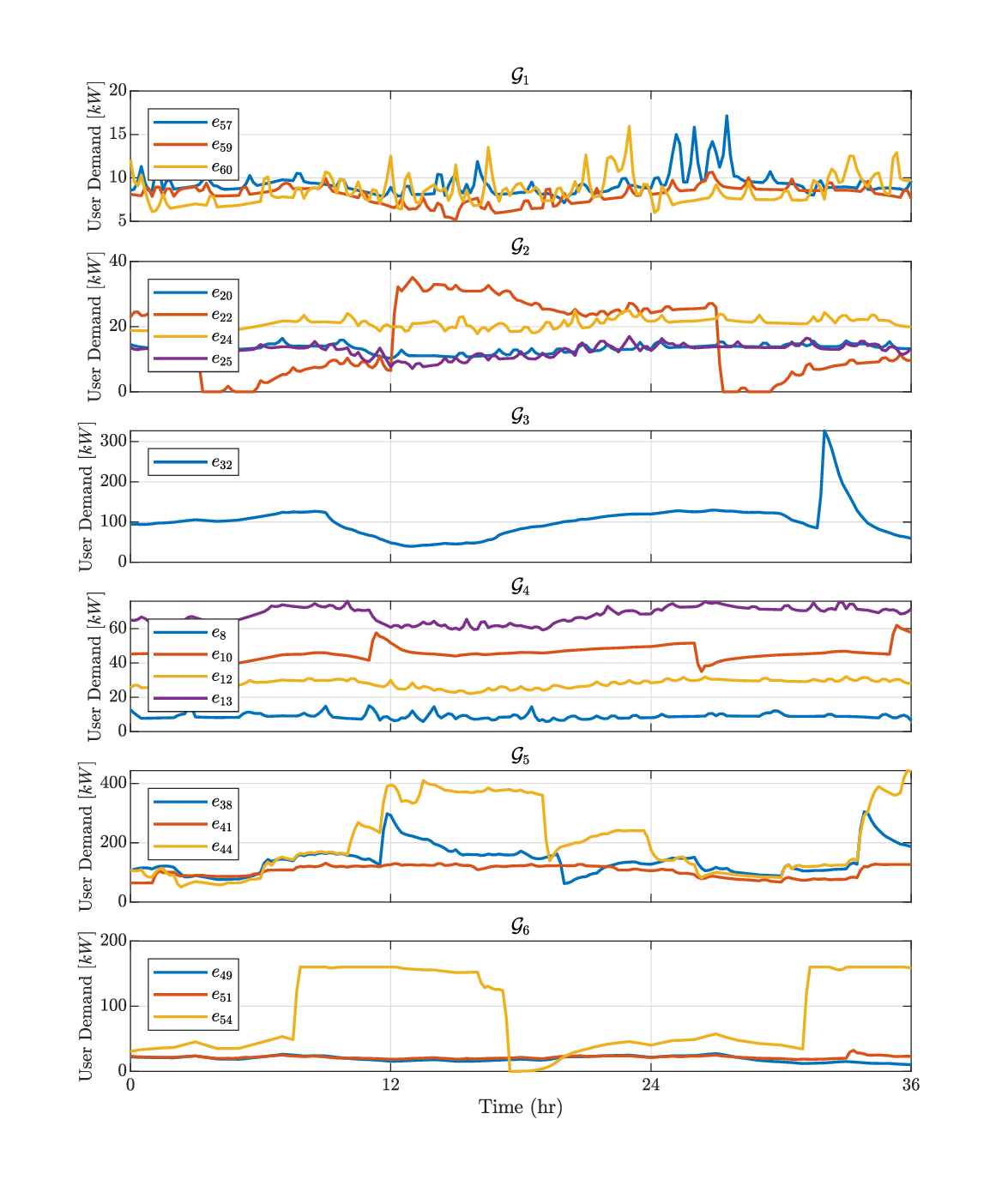}
    \caption{The nominal demands of the buildings in the DHN, organized by subgraph.}
    \label{fig:demand}
\end{figure}
\begin{figure}
    \centering
    \begin{subfigure}{\columnwidth}
        \centering
        \includegraphics[width=.7\linewidth,trim={5 0 30 5},clip]{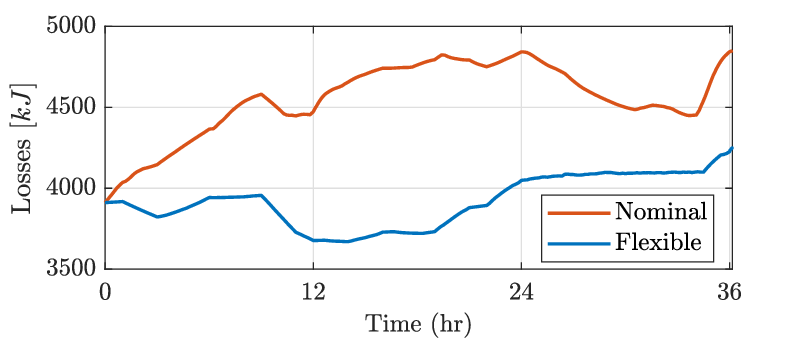}
        \caption{Network losses.}
        \label{fig:losses}
    \end{subfigure}
    \begin{subfigure}{\columnwidth}
        \centering
        \includegraphics[width=.7\linewidth,trim={25 0 35 5},clip]{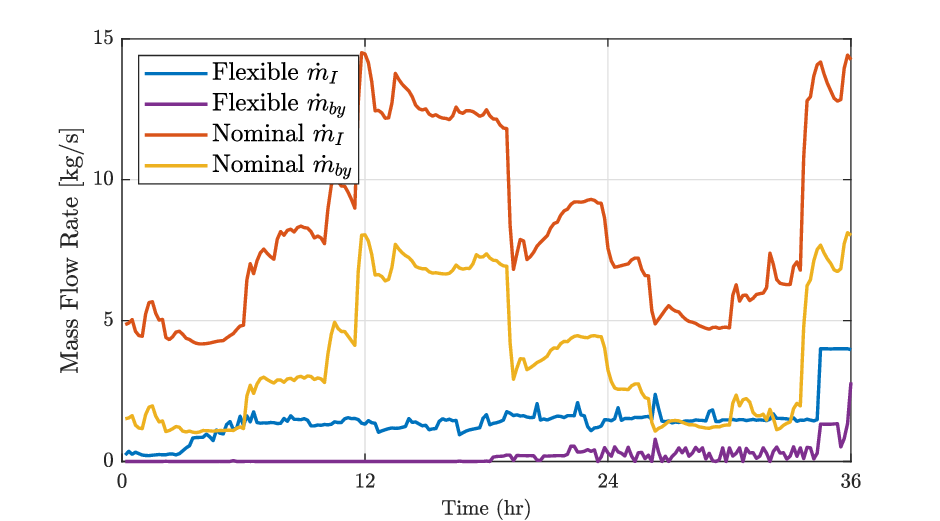}
        \caption{Plant and bypass mass flow rates.}
        \label{fig:mdot_tot}
    \end{subfigure}
    \begin{subfigure}{\columnwidth}
        \centering
        \includegraphics[width = .7\columnwidth,trim={15 0 30 5},clip]{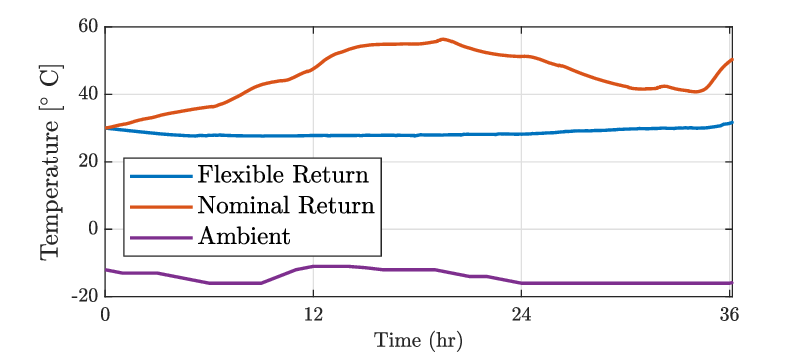}
        \caption{Return and ambient temperatures.}
        \label{fig:temp}
    \end{subfigure}
    \caption{Network level results comparing the optimized and nominal cases.}\label{fig:system_metrics}
\end{figure}

The case study considered in this paper is an 18-user network with realistic residential and commercial buildings and network parameters taken from literature. Full details of the network configuration and building characteristics can be found in \cite{blizardUsingFlexibilityEnvelopes2024b}. The supply temperature is assumed to be constant at $T_0 = 80\degree C$ the building temperature deviation was chosen as $\Delta T_b = 2\degree C$, and the ambient temperature profile is shown in \cref{fig:temp}. The nominal building demand is shown in \cref{fig:demand}. The minimum valve position $\theta_{min}= 0.01$ and the valve coefficient $\mu= 5.74$, based on the other pipes' friction coefficients. \par
The cost function chosen for \cref{eq:i_cost} was 
\begin{equation}
    {hA_s}_i\left(T_{{\mathcal{E}_N}_i}-T_{amb}\right)+ \frac{1}{\operatorname{card}\left(\mathcal{V}_{U_i}\right)}\left(\frac{SOC_{U_i}}{C_i\Delta T_b}\right)^2
\end{equation}
where the first term is the heat losses in the pipes and the second is the percent of used flexibility envelope normalized by the number of buildings in the subnetwork. This cost function was chosen to minimize losses while penalizing user discomfort.
Additionally, $\omega = 0.5$, and the values for epsilon were 
\begin{equation}
    \epsilon = \begin{bmatrix}
        0.1 & 1 & 1 & 0.2 & 0.2 & 0.5
    \end{bmatrix} ^T
\end{equation}
based on the relative magnitude of each variable type.
The 36-hour simulation was implemented in a receding horizon fashion where the control horizon was one hour with a sampling rate of 10 minutes. All results are compared to the nominal centralized solution where the buildings' flexibility envelopes were set to zero.\par
The results shown in \cref{fig:system_metrics} demonstrate that this controller greatly improves system performance over the nominal demand case. \Cref{fig:losses} shows that in the optimized case, there was a 14.0\% decrease in losses. \Cref{fig:mdot_tot} shows the overall decrease in mass flow rate needed to supply the network and the decrease in wasted mass flow sent through the bypass segments. While this drastic decrease in plant mass flow is not expected to persist after the buildings have used more of their state of energy, it is still indicative of the distributed control architecture's ability to find a good solution to approximate the centralized optimization problem. Finally, \cref{fig:temp} shows the average return temperature in the optimized and nominal cases. These results show a decrease of 37\% in average return temperature, caused by the diminished mass flow through the bypass segments, ensuring all returned water is at $T_{setR}$ in the optimized case.\par
\begin{figure}
    \centering
    \includegraphics[width=1\linewidth,trim={45 50 49 35},clip]{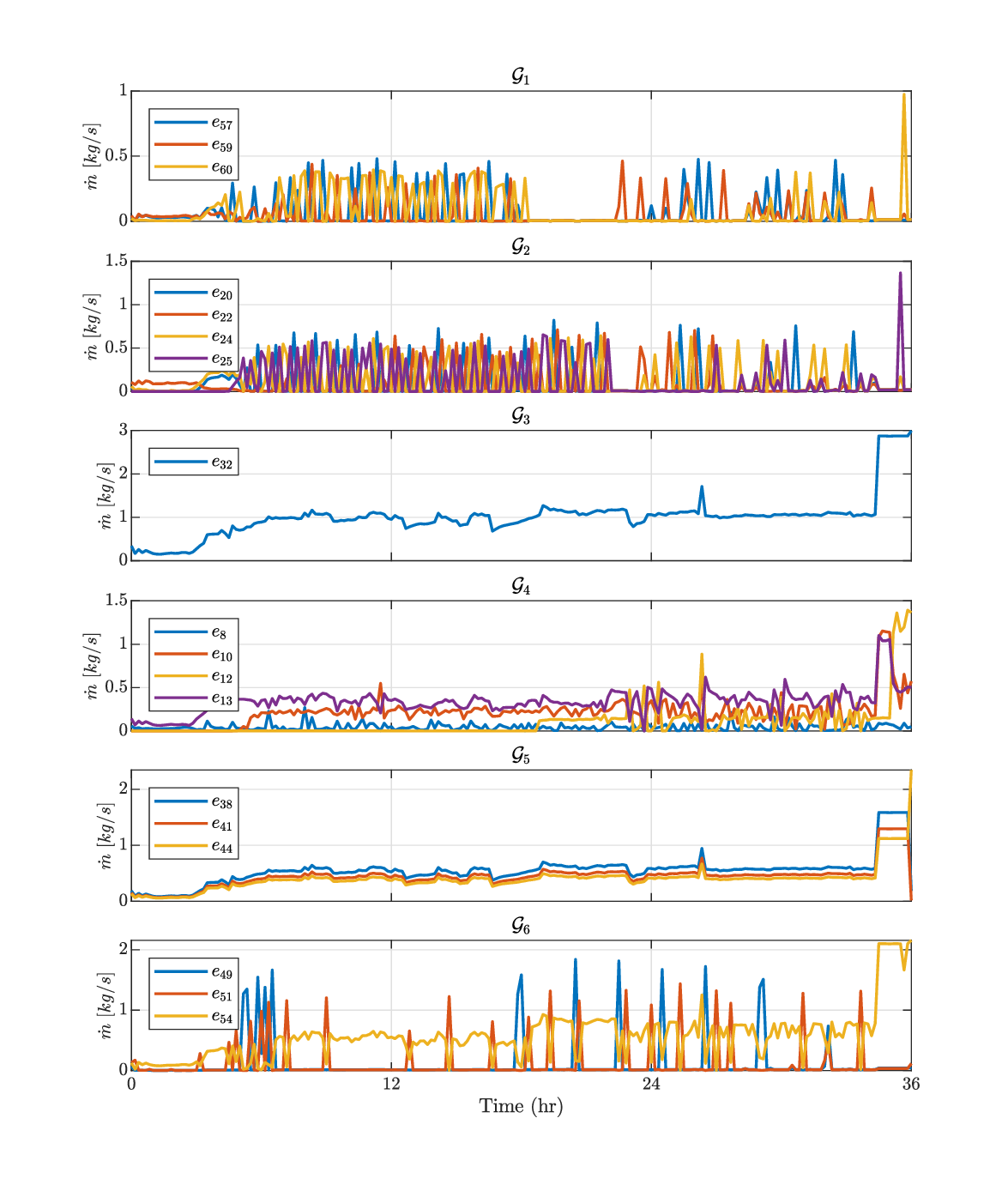}
    \caption{Mass flow rate delivered to users in optimized case.}
    \label{fig:mdot_user}
\end{figure}
\begin{figure}
    \centering
    \includegraphics[width=1\linewidth,trim={25 30 50 25},clip]{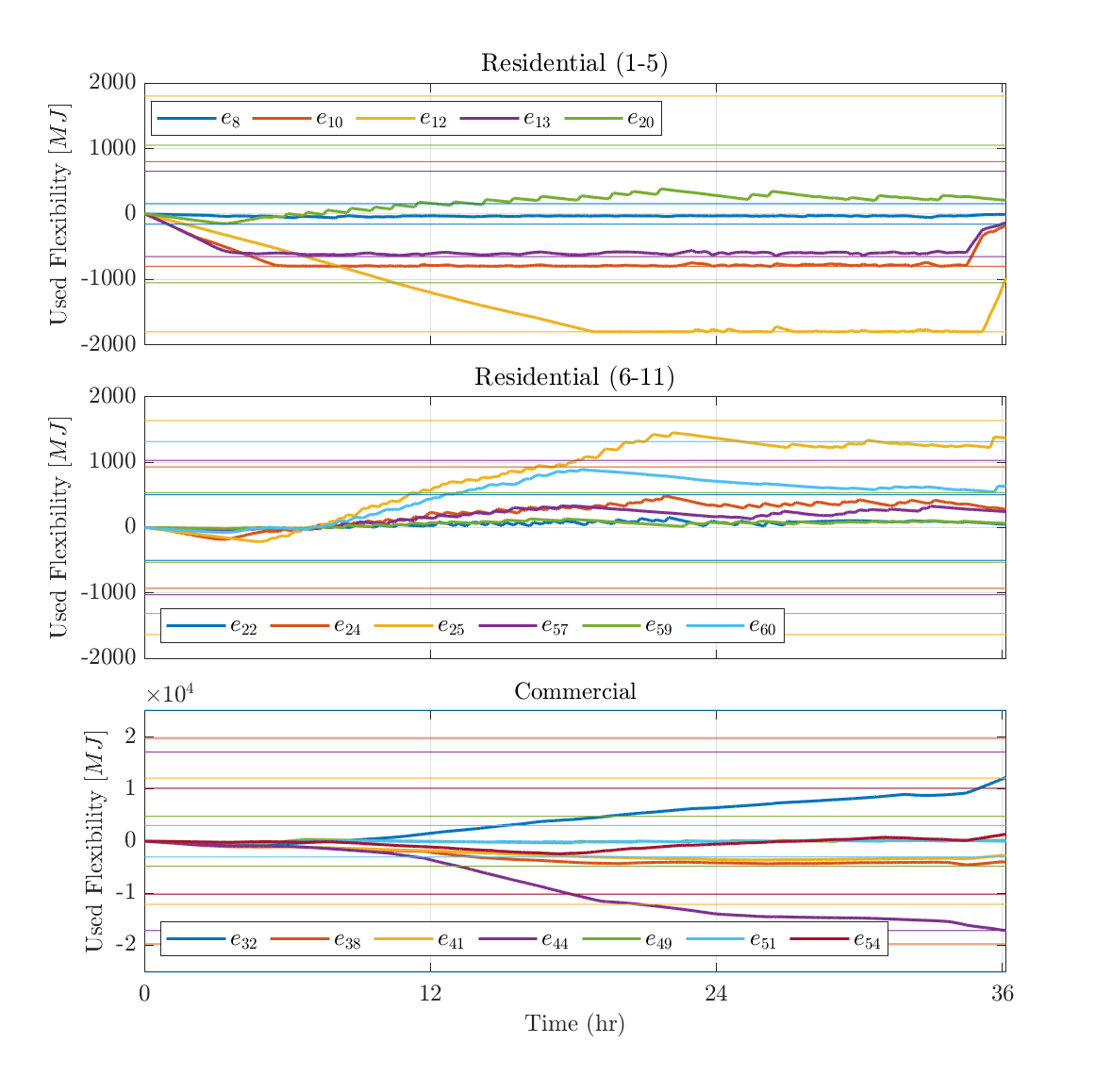}
    \caption{Flexibility used in optimized case.}
    \label{fig:usedflex}
\end{figure}

\Cref{fig:mdot_user,fig:usedflex} show the details of how the flow is distributed to the users and how this flow distribution affects the buildings' states of charge. Analyzing these results indicates how the optimized controller achieved the overall system performance improvements. In $\mathcal{G}_{1,2,4,6}$ the mass flow is traded off between users, minimizing wasted flow while ensuring the small flexibility envelopes of the connected users are never exceeded. $\mathcal{G}_5$, the plant pressure loss dictator,  chose to minimize the overall mass flow rate and deplete its buildings' state of energies, as it had no incentive to increase flow unless the restoration procedure was activated. $\mathcal{G}_3$ followed the same overall profile as $\mathcal{G}_5$ but chose to charge the single building in its network to avoid wasting flow. These results indicate that the distributed control framework effectively signals the subgraphs preferred behaviors without requiring the centralized optimization problem to be solved.

\section{Conclusion}
This work presents a communication-based dMPC scheme with an information passing scheme and a feasibility restoration procedure. The effectiveness of this control was demonstrated in a case study considering an 18-user DHN. Results showed the developed controller's ability to coordinate systems behaviors and improve performance, reducing losses by 14\% and return temperature by 37\%.\par
In this work, the systems were allowed to iterate until convergence, leading the calculation time of some iterations to exceed the control horizon of 10 minutes. Future work will look to address this issue by 1) using the properties of the developed controller to develop early stopping criteria and 2) improving convergence through improved partitioning of the system. Additionally, future work will look to extend the case study to longer simulations and eventually real DHNs.

\bibliographystyle{IEEEtran}
\bibliography{sources}

\begin{thebibliography}{10}
\providecommand{\url}[1]{#1}
\csname url@samestyle\endcsname
\providecommand{\newblock}{\relax}
\providecommand{\bibinfo}[2]{#2}
\providecommand{\BIBentrySTDinterwordspacing}{\spaceskip=0pt\relax}
\providecommand{\BIBentryALTinterwordstretchfactor}{4}
\providecommand{\BIBentryALTinterwordspacing}{\spaceskip=\fontdimen2\font plus
\BIBentryALTinterwordstretchfactor\fontdimen3\font minus \fontdimen4\font\relax}
\providecommand{\BIBforeignlanguage}[2]{{%
\expandafter\ifx\csname l@#1\endcsname\relax
\typeout{** WARNING: IEEEtran.bst: No hyphenation pattern has been}%
\typeout{** loaded for the language `#1'. Using the pattern for}%
\typeout{** the default language instead.}%
\else
\language=\csname l@#1\endcsname
\fi
#2}}
\providecommand{\BIBdecl}{\relax}
\BIBdecl

\bibitem{vandermeulenControllingDistrictHeating2018}
A.~Vandermeulen, B.~{van der Heijde}, and L.~Helsen, ``Controlling district heating and cooling networks to unlock flexibility: {{A}} review,'' \emph{Energy}, vol. 151, pp. 103--115, May 2018.

\bibitem{salettiEnablingSmartControl2021}
C.~Saletti, N.~Zimmerman, M.~Morini, K.~Kyprianidis, and A.~Gambarotta, ``Enabling smart control by optimally managing the {{State}} of {{Charge}} of district heating networks,'' \emph{Applied Energy}, vol. 283, p. 116286, Feb. 2021.

\bibitem{vanoevelenTestingEvaluationSmart2023}
T.~Van~Oevelen, T.~Neven, A.~Br{\`e}s, R.-R. Schmidt, and D.~Vanhoudt, ``Testing and evaluation of a smart controller for reducing peak loads and return temperatures in district heating networks,'' \emph{Smart Energy}, vol.~10, p. 100105, May 2023.

\bibitem{blizardUsingFlexibilityEnvelopes2024b}
A.~Blizard, C.~N. Jones, and S.~Stockar, ``Using {{Flexibility Envelopes}} for the {{Demand-Side Hierarchical Optimization}} of {{District Heating Networks}},'' in \emph{2024 {{European Control Conference}} ({{ECC}})}, Jun. 2024, pp. 273--278.

\bibitem{mullerEconomicDistributedModel2017}
M.~A. M\&uuml;ller and F.~Allg\&ouml;wer, ``Economic and {{Distributed Model Predictive Control}}: {{Recent Developments}} in {{Optimization-Based Control}},'' \emph{SICE Journal of Control, Measurement, and System Integration}, vol.~10, no.~2, pp. 39--52, Mar. 2017.

\bibitem{rostamiADMMbasedDistributedModel2017}
R.~Rostami, G.~Costantini, and D.~G{\"o}rges, ``{{ADMM-based}} distributed model predictive control: {{Primal}} and dual approaches,'' in \emph{2017 {{IEEE}} 56th {{Annual Conference}} on {{Decision}} and {{Control}} ({{CDC}})}, Dec. 2017, pp. 6598--6603.

\bibitem{houskaAugmentedLagrangianBased2016}
B.~Houska, J.~Frasch, and M.~Diehl, ``An {{Augmented Lagrangian Based Algorithm}} for {{Distributed NonConvex Optimization}},'' \emph{SIAM Journal on Optimization}, vol.~26, no.~2, pp. 1101--1127, Jan. 2016.

\bibitem{yaoStateArtReview2021}
Y.~Yao and D.~K. Shekhar, ``State of the art review on model predictive control ({{MPC}}) in {{Heating Ventilation}} and {{Air-conditioning}} ({{HVAC}}) field,'' \emph{Building and Environment}, vol. 200, p. 107952, Aug. 2021.

\bibitem{huModelPredictiveControl2021}
J.~Hu, Y.~Shan, J.~M. Guerrero, A.~Ioinovici, K.~W. Chan, and J.~Rodriguez, ``Model predictive control of microgrids -- {{An}} overview,'' \emph{Renewable and Sustainable Energy Reviews}, vol. 136, p. 110422, Feb. 2021.

\bibitem{blizardGraphBasedTechniqueAutomated2024}
A.~Blizard and S.~Stockar, ``A {{Graph-Based Technique}} for the {{Automated Control-Oriented Modeling}} of {{District Heating Networks}},'' \emph{Journal of Dynamic Systems, Measurement, and Control}, vol. 146, no. 041001, Mar. 2024.

\bibitem{reyndersGenericCharacterizationMethod2017}
G.~Reynders, J.~Diriken, and D.~Saelens, ``Generic characterization method for energy flexibility: {{Applied}} to structural thermal storage in residential buildings,'' \emph{Applied Energy}, vol. 198, pp. 192--202, Jul. 2017.

\bibitem{christofidesDistributedModelPredictive2013}
P.~D. Christofides, R.~Scattolini, D.~{Mu{\~n}oz de la Pe{\~n}a}, and J.~Liu, ``Distributed model predictive control: {{A}} tutorial review and future research directions,'' \emph{Computers \& Chemical Engineering}, vol.~51, pp. 21--41, Apr. 2013.

\end{thebibliography}

\end{document}